\documentclass[aps,prd,twocolumn,showpacs,preprintnumbers,
amsmath,amssymb,superscriptaddress,showkeys]{revtex4}
\usepackage{dcolumn}% Align table columns on decimal point
\usepackage{bm}% bold math

\begin{document}

\preprint{TIT/HEP--517}
\preprint{hep-th/0403009}
\title{Radius Stabilization 
in a Supersymmetric Warped Compactification}

\author{Minoru Eto}
\affiliation{Department of Physics, Tokyo Institute 
of Technology, Tokyo 152-8551, JAPAN}

\author{Nobuhito Maru}
\affiliation{\mbox{Theoretical Physics Laboratory, RIKEN, 
Saitama 351-0198, JAPAN} 
\\[2mm]
\mbox{\textnormal{\texttt{
meto@th.phys.titech.ac.jp,
maru@postman.riken.go.jp,
nsakai@th.phys.titech.ac.jp
}}}}

\author{Norisuke Sakai}
\affiliation{Department of Physics, Tokyo Institute of 
Technology, Tokyo 152-8551, JAPAN}

\date{\today}

\begin{abstract}
A supersymmetric (SUSY) model of radius stabilization 
is constructed for the $S^1/Z_2$ warped compactifications 
with a hypermultiplet in five dimensions. 
Requiring the continuity of scalar field across the 
boundaries, we obtain radius stabilization preserving 
SUSY, realizing the SUSY 
extension of the Goldberger-Wise mechanism. 
Even if we allow discontinuity of the $Z_2$ odd field 
across the boundary, we always obtain SUSY preservation 
but obtain the radius stabilization only when the 
discontinuity is fixed by other mechanisms. 
\end{abstract}


\maketitle
%%%%%%%%%%%%%%%%%%%%%%
\section{Introduction}
%%%%%%%%%%%%%%%%%%%%%%
Motivated by branes in string theories, 
models with extra dimensions 
\cite{HoravaWitten} 
have been proposed 
to offer another possible solution to 
the gauge hierarchy problem in recent years \cite{ADD,RS},  
instead of the well-studied models with 
supersymmetry (SUSY)  \cite{DGSW}. 
In these brane-world scenarios, 
the weak scale is derived from the four dimensional 
Planck scale 
through the large volume suppression \cite{ADD} or 
through the warp factor \cite{RS} without 
fine-tuning of parameters. 
Many of these models have the compactification radius 
as one of the arbitrary parameters of the model, 
namely a moduli which 
are not determined by the dynamics of the model. 
%The particle associated with the fluctuation of the 
%radius is called radion. 
For the scenarios to be relevant for nature, 
it is necessary to find the mechanism stabilizing 
the radius. 
For flat space models \cite{ADD}, a number of stabilization 
mechanisms have been proposed. 
An interesting possibility is to use the topological 
winding number as the origin of the stability 
\cite{MSSS}--\cite{SakaiSugisaka}. 
Explicit models with topological stability 
have been worked out in four-dimensional 
models with four SUSY in flat space 
\cite{MSSS}--\cite{SakaiSugisaka}. 
The model has also been successfully embedded 
into four-dimensional supergravity 
\cite{EMSS, EMS1} 
for warped compactifications  \cite{RS}. 
Models with the topological stability 
in five-dimensions 
with eight SUSY are being worked out \cite{EMS2}. 

On the other hand, 
one of the popular models of radius stabilization 
for warped compactifications \cite{RS} 
is the model of 
Goldberger and Wise 
which uses 
a bulk scalar field 
\cite{GW}. 
They introduced appropriate potentials on the brane 
to pin-down the values of the scalar field. 
Then the bulk dynamics of the scalar field 
generates a potential to stabilize 
the radius. 
They have studied the limit where the backreaction 
to the warped geometry can be neglected. 
Similar stabilization 
mechanisms have been much studied using more general 
scalar fields \cite{DFGK, SkTo, LMS}. 

SUSY is also quite useful in brane-world 
scenarios. 
The topological defects such as walls often 
break part of SUSY. 
Therefore the effective theories on the wall 
can possess half of SUSY of higher dimensional 
theories leading to the well-studied ${\cal N}=1$ 
SUSY models in four-dimensions. 
SUSY also helps to obtain solutions of the topological 
defects needed for the brane-world scenarios, 
since the BPS equations for the partial SUSY conservation 
are much easier to solve. 
Even the warped compactification models using orbifold 
\cite{RS} have been 
realized as a zero-width limit of the domain wall 
solutions in supergravity \cite{EMSS, EMS1}. 
Considering SUSY warped compactification models is 
also well motivated 
from the viewpoint of SUSY flavor problem \cite{AMSB}. 
Separation of the hidden and the visible sectors 
in extra dimensions 
forbids the contact interactions between these two sectors 
causing flavor-violating scalar masses by the 
higher dimensional locality. 
Therefore, it is interesting to examine whether the 
Goldberger-Wise mechanism 
can be extended to SUSY theories. 
A simple SUSY model of radius stabilization is recently proposed 
and its related SUSY breaking phenomenology is also discussed \cite{MO}.

The purpose of our paper is to 
propose a simple model of SUSY extention 
of the Goldberger-Wise mechanism of 
radius stabilization and to analyze the consequences. 
We find that 
SUSY is always preserved 
(four out of eight SUSY), with 
no additional contribution to the vacuum energy, 
justifying 
our assumption of neglecting backreaction to the 
background metric. 
We also find that the radius is stabilized 
as long as we insist on continuity across the 
orbifold fixed points for all the scalar fields 
including the $Z_2$ odd scalar field. 
If we allow a discontinuity of 
the $Z_2$ odd field at the boundary of $S^1/Z_2$ 
as a free parameter, 
we obtain solutions with single arbitrary parameter. 
Consequently the radius 
appears to become undetermined 
free parameters. 
We can understand the result by imagining 
a zero-width limit of a domain wall configuration 
\cite{EMSS, AFNS} 
which is made of the $Z_2$ odd scalar field. 
In the wall solution, the amount of the energy density 
generated by the wall is related to how rapidly 
the scalar field changes across the wall. 
In the zero-width limit, the discontinuities 
of the $Z_2$ odd field at 
both boundaries should be determined 
by the equations of motion for the scalar field 
if the appropriate dynamics is installed for the 
$Z_2$ odd scalar field to form the domain wall. 
Therefore we should consider the discontinuity to be 
a given parameter determined by (yet unspecified) 
wall dynamics. 
If we regard the discontinuity of the $Z_2$ odd field 
to be a given fixed input, the radius 
is uniquely determined.

In sec.2, we introduce our model and give solutions 
for generic situations with possible discontinuities 
for $Z_2$ odd scalar field. 
In sec.3, we give our results and discuss their 
physical implications.

%%%%%%%%%%%%%%%%%%%%%%%%%%%%%%%%%%%%%%%%%%%%%%%
\section{Our Model}
%%%%%%%%%%%%%%%%%%%%%%%%%%%%%%%%%%%%%%%%%%%%%%%
We consider a SUSY theory with a hypermultiplet 
in five dimensions 
compactified on an orbifold $S^1/Z_2$. 
The metric is given by \cite{RS} 
\begin{equation}
ds^2 = e^{-2r\sigma} \eta_{\mu\nu} dx^\mu dx^\nu 
+ r^2 dy^2, \sigma(y) = k|y|, 0 \le y \le \pi, 
\label{eq:RSmetric}
\end{equation}
where $\mu, \nu =0, \cdots, 3$, and  
$r$ is the $S^1$ 
compactification radius of the extra dimension 
$x^4\equiv y$. 
Since we assume the backreaction 
from the bulk scalar field 
is negligible \cite{GW}, the background 
warped (AdS) geometry (\ref{eq:RSmetric}) 
is fixed and the supergravity multiplet is 
treated as frozen and nondynamical. 
Using the four SUSY superfield formalism 
\cite{AGW, MSS, MirabelliPeskin, Hebecker, KakimotoSakai}, 
the Lagrangian is given by \cite{MP}
\begin{eqnarray}
{\cal L}_5 &=& \int d^4 \theta r e^{-2r\sigma}
\left[ |H|^2 + |H^c|^2 \right] \nonumber \\
&&+ \left[ \int d^2 \theta e^{-3r \sigma} 
\left\{
H^c (\partial_y - (\frac{3}{2} - c) r \sigma')H \right. \right. \nonumber \\
&& \left.\left. + \delta(y) W_0 + \delta(y-\pi) W_\pi \right\} + {\rm h.c.}
\right], 
\end{eqnarray}
where the prime denotes the differentiation 
with respect to $y$, and the constant $c$ 
specifies the bulk mass of hypermultiplets. 
The chiral 
scalar superfields $H$ and $H^c$ 
in the four SUSY 
notation form a 
hypermultiplet of eight SUSY. 
Since eight SUSY does not allow 
superpotentials among hypermultiplets in the bulk, 
we can introduce the  superpotentials only 
on the boundaries  $y=0$, $\pi$
(orbifold fixed points) which are denoted as 
$W_{0}$, $W_{\pi}$, respectively. 
The orbifolding on $S^1/Z_2$ breaks the eight SUSY 
maintaining only four SUSY. 
We shall assign even $Z_2$ parity to 
$H$, odd $Z_2$ parity to $H^c$, respectively. 
Since only the even field can have nonvanishing values 
on the boundaries, 
the boundary superpotential can have only parity 
even field $H$.

For simplicity, we shall take the quadratic 
boundary superpotential with a unique SUSY vacuum 
\begin{eqnarray}
\label{bdrsp}
W_{0,\pi} = \frac{1}{2}H^2 - v_{0,\pi} H, 
\end{eqnarray}
where $v_{0,\pi}$ are constants with mass dimension 3/2. 
It is useful to make a rescaling 
\begin{eqnarray}
(H, H^c) \to e^{r \sigma} (H, H^c). 
\end{eqnarray}
Then the Lagrangian becomes 
\begin{eqnarray}
\label{rescaled}
{\cal L}_5 &\to& 
\int d^4 \theta r(|H|^2 + |H^c|^2) \nonumber \\
&&+ \left[ \int d^2 \theta 
\left\{ e^{-r \sigma} H^c \partial_y H
+ e^{-r \sigma} \left( c - \frac{1}{2} \right) 
r \sigma' H^c H \right. \right. \nonumber \\
&& \left. \left. + \delta(y) 
\left( \frac{1}{2} 
e^{-r\sigma} H^2 - v_0 e^{-2r \sigma}H \right) \right. \right. \nonumber \\
&& \left. \left. + \delta(y-\pi) 
\left( \frac{1}{2} 
e^{-r\sigma} H^2 - v_\pi e^{-2r \sigma}H \right) 
\right\} + {\rm h.c.} \right]. 
\label{rescaled2}
\end{eqnarray}
It is straightforward to derive 
the auxiliary fields part of 
Lagrangian, 
\begin{eqnarray}
\label{aux1}
{\cal L}_{{\rm aux}} 
&=&  
r(|F|^2 + |F^c|^2) \nonumber \\
&&+ \left[ 
e^{-r\sigma}
\left\{
F^c \partial_y H 
+ r\sigma' H^c F 
- \partial_y H^c F \right. \right. \nonumber \\
&& \left. \left. + \left( c - \frac{1}{2} \right) r \sigma' (F^c H + H^c F) 
\right\} \right. \nonumber \\
&& \left. +\delta(y) 
(e^{-r\sigma}H - v_0 e^{-2r \sigma})F \right. \nonumber \\
&& \left. +\delta(y-\pi) 
(e^{-r\sigma}H - v_\pi e^{-2r \sigma})F 
+ {\rm h.c.}
\right] \\
&=& 
\label{eq:AuxLag}
- r(|F|^2 + |F^c|^2). 
\end{eqnarray}
In the second equality, we used the 
 equations of motion (EOM) 
for auxiliary fields derived from Eq.~(\ref{aux1}) 
\begin{eqnarray}
\label{F1}
F^* &\!\!\!=&\!\!\! 
\frac{e^{-r \sigma}}{r} 
\left[
\partial_y H^c - \left( c + \frac{1}{2} \right) r \sigma' H^c 
- e^{r \sigma} W_b
\right] \nonumber \\
 &\!\!\!=&\!\!\! 
\frac{e^{-r \sigma}}{r} 
\left[e^{(c+\frac{1}{2})r \sigma} 
\partial_y \left(e^{-(c+\frac{1}{2})r \sigma}
H^c \right)
- e^{r \sigma} W_b
\right], \\
\label{Fc*}
F^{c*} &\!\!\!=&\!\!\! -\frac{e^{-r \sigma}}{r} 
\left[
\partial_y H + \left( c - \frac{1}{2} \right) r \sigma' H 
\right], \nonumber \\
&\!\!\!=&\!\!\!
-\frac{e^{-r \sigma}}{r} 
\left[e^{-(c-\frac{1}{2})r \sigma}
\partial_y \left(e^{(c-\frac{1}{2})r \sigma}
H \right)
\right], 
\end{eqnarray}
where 
\begin{eqnarray}
W_b &=& \delta(y)
\left( e^{-r\sigma}H - v_0 e^{-2r \sigma} \right) \nonumber \\
&&+ \delta(y-\pi) 
\left( e^{-r\sigma}H - v_\pi e^{-2r \sigma} \right) \\
&\equiv& \delta(y)\tilde{W}_0 + \delta(y-\pi) \tilde{W}_\pi. 
\end{eqnarray}
Here and the following, we shall use the same notation 
for scalar fields as the superfields. 
It is important to notice that the auxiliary field $F$ 
in Eq.(\ref{F1}) contains delta function in general 
which introduces singular interaction terms like 
$(\delta(y))^2$ as noted previously \cite{MirabelliPeskin}. 

In conformity with the warped metric compactifications in 
Eq.(\ref{eq:RSmetric}), we are interested in 
%wish to obtain solutions of the equations of motion for 
the configurations of the scalar fields 
$H, H^c$ as functions of extra dimensional 
coordinate $y$ only. 
The scalar field $H$ with the even $Z_2$ parity 
does not vanish at the boundaries. 
However, the scalar field $H^c(y)$ with odd $Z_2$ parity 
has to vanish at the boundaries. 
To make this point clear, we rewrite the 
parity odd field $H^c(y)$ in terms of 
%the sign function $\hat \varepsilon (y)$ and 
a parity even field $h^c(y)$ as 
\begin{equation}
H^c(y) = \hat \varepsilon(y)h^c(y), 
\end{equation}
where $\hat\varepsilon(y)$ is a sign function of $y$ 
\begin{equation}
\hat\varepsilon(y) \equiv 
\left\{
\begin{array}{ll}
-1,
\qquad&{\rm for}\ -\pi< y<0,\\
0,
\qquad&{\rm for}\ y=0, \pm \pi, \\
1, \qquad&{\rm for}\ 0<y<\pi .
\end{array}
\right.
\end{equation}
From physical grounds, we should consider field 
configurations of the physical scalar fields 
$H(y), H^c(y)$ which are continuous across the boundaries. 
This implies that we need to require 
\begin{equation}
h^c(0) = 
h^c(\pi) = 0. 
\end{equation}
In special circumstances like the zero-width limit of 
domain wall configurations, the odd $Z_2$ parity field 
$H^c$ can have a discontinuity across the boundaries. 
In order to examine such a general situation later, 
we shall temporarily allow discontinuities across 
the boundaries for $H^c$, corresponding to a finite 
nonvanishing values of $h^c(0)$ and $h^c(\pi)$. 
Apart from this subtlety, $h^c(y)$ is assumed 
to be a continuous~\cite{ft:disc} 
 and parity even function of $y$. 
Then, Eq.~(\ref{F1}) can be rewritten as 
\begin{equation}
F^* = \frac{e^{-r \sigma}}{r} 
\left[
e^{(c+\frac{1}{2})r\sigma} 
\partial_y (e^{-(c+\frac{1}{2})r\sigma}
\hat \varepsilon(y)h^c(y))
- e^{r \sigma} W_b
\right].
\label{F1'}
\end{equation}

As far as $H, H^c$ depend on $y$ only, 
it is sufficient to consider only the auxiliary fields 
part of the Lagrangian (\ref{eq:AuxLag}). 
Then the EOM for $H^c$ is given by
\begin{eqnarray}
0 &=& {\partial {\cal L} \over \partial H^{c*}}
-\partial_y{\partial {\cal L} \over 
\partial \partial_y H^{c*}}
=
\frac{1}{r}e^{-(c+\frac{1}{2})r\sigma}\partial_y 
\left[
e^{(c-\frac{3}{2})r\sigma} \right. \nonumber \\
&\times& \left. \left\{
e^{(c+\frac{1}{2})r\sigma}
\partial_y(e^{-(c+\frac{1}{2})r\sigma} 
\hat \varepsilon(y)h^c(y))
-e^{r\sigma}W_b
\right\}
\right]. 
\label{Hceom}
\end{eqnarray}
The singular part of the equations of motion 
(\ref{Hceom}) contains 
$\partial_y \delta(y),\partial_y \delta(y-\pi)$ 
originating from 
$\frac{\partial W_b}{\partial y}$ 
and $\partial_y^2 \hat \varepsilon(y)$ 
and reads 
\begin{eqnarray}
0 &=& \frac{e^{-2r\sigma}}{r}
\left[
\partial_y \delta(y)(2h^c(0) - \tilde{W}_0) \right. \nonumber \\
&& \left. + \partial_y \delta(y-\pi)(-2h^c(\pi) - e^{rk\pi} \tilde{W}_\pi) 
\right]. 
\end{eqnarray}
Thus we obtain the following boundary conditions
\begin{eqnarray}
\label{bchc0}
2h^c(0) &=& \tilde{W}_0 = H(0) - v_0, \\
\label{bchcpi}
-2h^c(\pi) &=& e^{rk\pi} \tilde{W}_\pi = H(\pi) - e^{-rk\pi}v_\pi. 
\end{eqnarray}
Taking these boundary conditions into account, 
we can immediately see that the auxiliary field 
$F$ contains no delta functions, 
\begin{equation}
F^* = 
\hat \varepsilon(y) \frac{e^{(c-\frac{1}{2})r\sigma}}{r} 
\partial_y (e^{-(c+\frac{1}{2})r\sigma}h^c). 
\label{regularF}
\end{equation}
It is interesting to observe that 
the pining of the scalar field values $H(0), H(\pi)$ at 
boundaries arises in order to satisfy the 
EOM at the boundary, resulting in no singular terms 
in the auxiliary field as a result. 
This is in contrast to the Goldberger-Wise 
model without SUSY which requires a sharp 
potential well by tuning coupling parameters. 
Returning to the remaining EOM of $h^c$
\begin{equation}
0=\frac{1}{r}e^{-(c+\frac{1}{2})r\sigma}
\partial_y 
\left\{
\hat \varepsilon(y) 
e^{(2c-1)r\sigma} \partial_y (e^{-(c+\frac{1}{2})r\sigma}h^c)
\right\}, 
\end{equation}
we find the general solution 
\begin{equation}
h^c(y) =
\left\{
\begin{array}{ll}
-\frac{C_1}{(2c-1)rk}e^{-(c-\frac{3}{2})r\sigma} 
+ C_2 e^{(c+\frac{1}{2})r\sigma},
\qquad&{\rm for}\ c \ne \frac{1}{2},\\
(C_1 |y| + C_2)e^{r\sigma},
\qquad&{\rm for}\ c = \frac{1}{2}, 
\end{array}
\right.
\label{Hcsol}
\end{equation}
where $C_{1,2}$ are integration constants. 
The auxiliary field $F(y)$ is given by 
\begin{equation}
F^*(y) = 
C_1 (\hat \varepsilon(y))^2 
\frac{e^{-(c-\frac{1}{2})r\sigma}}{r}
\label{Fsol}
\end{equation}
irrespective of $c$. 
Note that $F$ can be non-zero only in the bulk and 
trivially vanishes at the boundaries, because of 
$\hat \varepsilon(y=0,\pi)=0$, except when it is multiplied 
by singular functions like delta functions.

Let us turn to the EOM for $H$, 
\begin{eqnarray}
0 &=& \frac{1}{r}
\left[
e^{(c-\frac{1}{2})r\sigma}\partial_y
\left\{
e^{-(2c+1)r\sigma}\partial_y(e^{(c-\frac{1}{2})r\sigma}H)
\right\} \right. \nonumber \\
&& \left. +r F^* \left\{ \delta(y) 
+ e^{-r\sigma} \delta(y-\pi) \right\}
\right]. 
\label{Heom}
\end{eqnarray}
Since $F^*$ in Eq.~(\ref{Heom}) 
is multiplied by a delta function, 
it gives nonvanishing contributions at the 
fixed points $y=0,\pi$ 
in spite of 
the square of the sign function 
%$(\hat \varepsilon(y))^2$ 
in Eq.~(\ref{Fsol}) \cite{ABN}, 
\cite{MNO} : 
\begin{eqnarray}
\left(\varepsilon(y)\right)^{2n}
\delta(y) 
&=&
{1 \over 2n+1}\delta(y), 
\\ 
\left(\varepsilon(y)\right)^{2n+1}
\delta(y) 
&=&0, \qquad n = 0, 1, 2, \cdots. 
\nonumber 
\end{eqnarray} 
Then, the remaining EOM for $H$ reads 
\begin{eqnarray}
0 &=& 
e^{(c-\frac{1}{2})r\sigma}\partial_y
\left\{
e^{-(2c+1)r\sigma}\partial_y(e^{(c-\frac{1}{2})r\sigma}H)
\right\}
\nonumber \\
&+&{C_1 \over 3}
\left\{
%e^{-(c-\frac{1}{2})r\sigma}
\delta(y)
+e^{(c+\frac{1}{2})r
%\sigma
k\pi}\delta(y-\pi)
\right\}
. 
\label{regularHeom}
\end{eqnarray}
The EOM in the bulk ($y\not=0, \pi$) becomes 
\begin{equation}
e^{-(2c+1)r\sigma}\partial_y(e^{(c-\frac{1}{2})r\sigma}H) 
= C_3 \hat \varepsilon(y), 
%~(C_3:{\rm const})
\label{eq:SolBulk}
\end{equation}
with an integration constant $C_3$. 
The $Z_2$ parity transformation property 
requires the sign function $\hat \varepsilon(y)$ 
in the right hand. 
However, we still need to examine the equations of motion 
(\ref{regularHeom}) at the boundaries. 
The solution in the bulk (\ref{eq:SolBulk}) gives 
delta functions at the boundaries 
\begin{equation}
\partial_y( e^{-(2c+1)r\sigma}
\partial_y(e^{(c-\frac{1}{2})r\sigma}H)) 
= 2C_3 (\delta(y) - \delta(y-\pi)). 
\end{equation}
The equations of motion (\ref{regularHeom}) 
is satisfied at the boundaries only when these 
delta functions are cancelled each other : 
the delta function at $y=0$ cancells if 
\begin{eqnarray}
0 &=& 
%\delta(y)\left\{
2 C_3 + \frac{C_1}{3}
%\right\}
\end{eqnarray}
the delta function at $y=\pi$ cancells if 
\begin{eqnarray}
%+\delta(y-\pi)\left\{
0=-2 C_3e^{-(c-\frac{1}{2})k\pi}
 + \frac{C_1}{3}e^{(c+\frac{1}{2})rk\pi}
%\right\}
. 
\end{eqnarray}
These two conditions together imply 
\begin{equation}
%$
C_1=C_3=0
%$
. 
\end{equation}
The solution of 
Eq.(\ref{eq:SolBulk}) with 
$C_3=0$ is 
given by 
\begin{equation}
H(y) = C_4 e^{-(c-\frac{1}{2})r\sigma}. 
\label{Hsol}
\end{equation}
The solution of the 
Eqs.~(\ref{Fc*}) and (\ref{eq:SolBulk}) with 
$C_3=0$ gives 
%the auxiliary field $F^{c}$ as 
\begin{equation}
F^{c*}(y)=0. 
\end{equation}
The result $C_1=0$ also implies the vanishing auxiliary field 
$F$ in Eq.(\ref{Fsol}) 
\begin{equation}
F(y)=0, 
\end{equation}
and the $Z_2$ odd scalar field 
%$H^c(y)\equiv \hat\varepsilon (y) h^c(y)$ 
in Eq.(\ref{Hcsol}) as 
\begin{equation}
H^c(y)= \hat\varepsilon (y) h^c(y)=
 \hat\varepsilon (y) 
C_2 e^{(c+\frac{1}{2})r\sigma}
. 
\label{eq:Hc}
\end{equation}
Therefore we find that both auxiliary fields $F$ and $F^c$ 
vanish, and that SUSY is always preserved. 

Using Eqs.(\ref{Hsol}) and (\ref{eq:Hc}), 
we can determine 
the remaining two integration constants $C_{2}, C_{4}$ 
from the boundary conditions (\ref{bchc0}) and 
(\ref{bchcpi}) as  
\begin{eqnarray}
\label{C2}
C_2 &=& 
\frac{- v_0 e^{-2crk\pi} + v_\pi e^{-(c+\frac{3}{2})rk\pi}}
{2(1+e^{-2crk\pi})}
, \\
C_4 &=& 
\frac{
 v_0 e^{crk\pi} + v_\pi e^{-\frac{3}{2}rk\pi}
}{1+e^{-2crk\pi}}
. 
\label{C4}
\end{eqnarray}

%%%%%%%%%%%%%%%%%%%%%%%%%%%%%%%%%%%%%%%%%%%%%%%
\section{Results}
%%%%%%%%%%%%%%%%%%%%%%%%%%%%%%%%%%%%%%%%%%%%%%%
Since 
the auxiliary fields $F, F^c$ 
vanish, 
SUSY is always preserved. 
As we stated in the previous section, 
we primarily consider that the physical scalar fields 
$H$, $H^c$ should be continuous across the boundaries 
$y=0, \pi$. 
This implies that the field $h^c(y)$ should vanish 
at the boundaries : $h^c(0)=h^c(\pi)=0$. 
Then the 
boundary conditions (\ref{bchc0}) and (\ref{bchcpi}) 
force the 
boundary values $H(0)$ and $H(\pi)$ of $Z_2$ even field 
$H(y)$ to settle at the minimum of the boundary 
superpotential 
\begin{equation}
H(0) = v_0, 
\qquad 
H(\pi) = v_\pi e^{-rk\pi}
, 
\label{eq:GWBoundaryValue}
\end{equation}
as in the non-SUSY case \cite{GW}. 
Combining Eqs.~(\ref{Hsol}) and 
(\ref{eq:GWBoundaryValue}), we obtain 
\begin{equation}
v_0=v_\pi e^{(c-\frac{3}{2})rk\pi}, 
\label{eq:radius}
\end{equation}
which determines the radius. 
The radius stabilization is thus achieved 
in our SUSY model. 
Moreover, the vanishing contribution to the 
vacuum energy justifies 
our assumption of no backreaction to Einstein 
equation for the metric. 
These result precisely realizes the objective of 
the Goldberger-Wise model. 
In our SUSY model, however, the field values 
$H(0), H(\pi)$ are fixed at the boundaries 
by their EOM 
without tuning parameters of the potential 
in contrast to the non-SUSY Goldberger-Wise model. 
We also find that $C_2=0$ from Eq.(\ref{eq:Hc}) 
with $h^c(0)=h^c(\pi)=0$, and that 
\begin{equation}
h^c(y) = 0. 
\end{equation}

On the other hand, 
if we allow nonvanishing discontinuities 
of the $Z_2$ odd field $H^c(y)$ across the boundaries, 
we obtain more complications. 
Even in this case, the equations of motion requires 
these discontinuities $h^c(0),h^c(\pi)$ to be related 
as given in Eqs.(\ref{eq:Hc}) and (\ref{C2}) :
\begin{equation}
 h^c(\pi)= h^c(0) e^{(c+\frac{1}{2})rk\pi}.
 \label{eq:SUSYdisc}
\end{equation}
Then the solution in the previous section 
contains single arbitrary parameter, say $h^c(0)$ 
undetermined. 
Therefore the radius is determined only by fixing the 
discontinuity $h^c(0)$. 
In order to understand the physical significance of 
these discontinuities, it should be useful 
to consider the zero-width limit of 
domain wall solutions with a scalar field \cite{EMSS, AFNS}. 
Since a domain wall consists of a kink of scalar field 
in the extra dimension, the scalar field usually changes 
sign at the boundaries. This is precisely 
a feature of the $Z_2$ odd scalar field. 
Therefore it is tempting to identify the $Z_2$ odd 
scalar field with the scalar field forming the wall. 
In the zero-width limit, the wall energy is 
concentrated at the boundary as a delta-function 
and should be 
related to the discontinuity of the 
scalar field across the boundary. 
Therefore we believe that the amount of the discontinuity 
across the boundary $h^c(0)$ (and $h^c(\pi)$) should 
be determined by the yet unspecified dynamics to form 
the domain wall. 
Provided such a microscopic description is given, 
this discontinuity is a fixed input parameter 
in our situation. 
Then the radius of compactification is determined 
using the fixed parameter as the input. 

It may be instructive to compare our model with 
a model admitting an exact two wall solution stabilized 
by a winding number \cite{MSSS,MSSS2}, 
which was embedded into supergravity in four dimensions 
\cite{EMSS}. 
In this model with winding number, 
a chiral scalar field serves as a 
$Z_2$ odd field to form domain walls with the symmetry 
$S^1/Z_2$. 
If the width of the wall is finite, the two wall 
configuration is found to be non-BPS (SUSY 
is completely broken) and the radius is stabilized 
\cite{EMS1}. 
In the limit of vanishing width (keeping 
wall tension fixed), however, 
the $Z_2$ odd scalar field of this model has 
no discontinuities across the orbifold boundaries 
leaving only boundary vacuum energy as a remnant. 
Then the model 
reduces to the 
Randall-Sundrum model \cite{RS}, and 
the scalar field 
is frozen 
in the zero-width limit without any other field available 
for the Goldberger-Wise 
type mechanism of radius stabilization to work. 
In fact, the 
two-wall solution 
can be regarded as a BPS configuration 
preserving half of the bulk SUSY \cite{ABN, FLP}, 
and the radius is undetermined in the zero width limit 
\cite{EMSS}. 
The scalar field forming the wall 
acts as a stabilizer field only for finite 
width of the wall, with fully broken SUSY. 
On the contrary, our present SUSY model of radius 
stabilization has the $Z_2$ even field $H$ and 
the boundary superpotential, which provide 
the stabilization mechanism 
%of the compactification radius $r$ 
preserving SUSY 
(assuming continuity of fields).

The advantages of our model are as follows. 
First, the stabilization of radius is maintained perturbatively, 
since the stabilization condition is determined 
by the F-flatness conditions. 
Even if the corrections to K\"ahler potential are considered, 
the conditions remain unchanged as long as 
the K\"ahler metric is non-singular and positive definite after 
quantum corrections. 
Second, we do not necessarily need to tune the warp factor to be 
$e^{-rk \pi} \sim 10^{-16}$ since the hierarchy problem can be 
solved by SUSY preserved on the boundaries. 
This fact offers more possibilities for the viable model 
construction. 
Third, as we mentioned in Introduction, 
the radius stabilization in SUSY models are required to address 
the SUSY flavor problem in the context of the brane world. 
Supersymmetric radius stabilization is phenomenologically favored 
as discussed in \cite{MO}.

Finally, we comment on the difference between our model and 
the model in \cite{MO}. 
In the model of \cite{MO}, 
there are always the discontinuities of the $Z_2$ odd scalar field 
across the boundaries because of the boundary superpotential linear in 
$Z_2$ even chiral superfield $H$. 
But in our model, the case without discontinuities is also allowed 
as mentioned in the text.

In summary, 
we have proposed a simple model of stabilizing 
the compactification radius 
in SUSY warped compactifications 
%Randall-Sundrum model 
with a hypermultiplet. 
By solving the equations of motion, we find that 
SUSY is always preserved. 
If the $Z_2$ odd scalar field of the 
hypermultiplet has no discontinuities 
across the boundaries, the $Z_2$ even scalar field 
settles at the minimum of the boundary superpotential, 
and the radius is determined 
by Eq.~(\ref{eq:radius}). 
This corresponds to a SUSY version of the 
Goldberger-Wise model. 
More generally, if we allow discontinuities of the 
 $Z_2$ odd scalar field across the boundaries, 
the $Z_2$ even scalar field 
does not necessarily settle at the minimum of the 
boundary superpotential, and the radius is stabilized 
only after fixing the discontinuity at the boundary 
by using 
the yet unspecified dynamics of forming the 
domain wall.

%%%%%%%%%%%%%%%%%%%%%%%%%%%%%%%%%%%%%%%%%%%%%%%%%%%%%%%%%%%%%%%%%%%%%%%%%%%%%%%
\section*{Acknowledgements}
One of the authors (M.E.) gratefully acknowledges 
support from the Iwanami Fujukai Foundation and 
from a 21st Century COE Program at 
Tokyo Tech "Nanometer-Scale Quantum Physics" by the 
Ministry of Education, Culture, Sports, Science 
and Technology.
This work is supported in part by Grant-in-Aid for Scientific 
Research from the Ministry of Education, Culture, Sports, 
Science and Technology, Japan No.13640269 and 
No.16028203 for the priority area ``Origin of mass'' 
(NS) and by Special Postdoctoral Researchers Program 
at RIKEN (NM). 
%%%%%%%%%%%%%%%%%%%%%%%%%%%%%%%%%%%%%%%%%%%%%%%%%%%%%%%%%%%%%%%%%%%%%%%%%%%%%%%%

%\appendix


\begin{thebibliography}{100}
\bibitem{ft:disc}
  First derivative may be discontinuous, as seen in 
  the solution in Eq.~(\ref{Hcsol}).
\end{thebibliography}
\end{document}